\documentclass[twocolumn,prb,amsmath,amssymb,showpacs]{revtex4}

\usepackage{graphicx}

\usepackage{dcolumn}

\usepackage{bm}

\newcommand{\gma}{$\rm{Ga_{1-x}Mn_{x}As}$}

\newcommand{\tc}{$T_{\rm{C}}$}
\newcommand{\tgma}{$t_{\rm{gma}}$}
\newcommand{\tspacer}{$t_{\rm{spacer}}$}
\newcommand{\he}{$H_{\rm{e}}$}
\newcommand{\hc}{$H_{\rm{c}}$}

\begin{document}
\title{Spin Valve Effect in Self-exchange Biased Ferromagnetic Metal/Semiconductor Bilayers}
\author{M.\ Zhu}
\author{M.\ J.\ Wilson}
\author{B.\ L.\ Sheu}
\author{P.\ Mitra}
\author{P.\ Schiffer}
\author{N.\ Samarth}\email{nsamarth@psu.edu}
\affiliation{Dept. of Physics and Materials Research Institute, The Pennsylvania State University, University Park PA 16802}

\begin{abstract}
We report magnetization and magetoresistance measurements in hybrid ferromagnetic metal/semiconductor heterostructures comprised of MnAs/(Ga,Mn)As bilayers. Our measurements show that the (metallic) MnAs and (semiconducting) (Ga,Mn)As layers are exchange coupled, resulting in an exchange biasing of the magnetically softer(Ga,Mn)As layer that weakens with layer thickness. Magnetoresistance measurements in the current-perpendicular-to-the-plane geometry show a spin valve effect in these self-exchange biased bilayers. Similar measurements in MnAs/p-GaAs/(Ga,Mn)As trilayers show that the exchange coupling diminishes with spatial separation between the layers.
\end{abstract}
\maketitle

Fundamental studies of spin-dependent transport and exchange coupling in ferromagnetic (FM) multilayers have played a central role in spintronics,\cite{Wolf:2001ev} with key emphasis on phenomena such as tunneling magnetoresistance, the spin valve effect,\cite{Tsymbal:2001lr} and exchange biasing.\cite{Nogues:1999uq} Within the particular context of semiconductor-based spintronics,\cite{Awschalom:2007qy,Samarth:2004rx} such studies have centered on heterostructures derived from the ferromagnetic semiconductor (Ga,Mn)As,\cite{Macdonald:2005rx,Jungwirth:2006fj} aiming at FMS versions of conventional metal spintronics phenomena such as the spin valve effect in trilayers,\cite{Xiang:2007prb} exchange biasing by an antiferromagnetic (AFM) pinning layer,\cite{Eid:2004rx,Dziatkowski:2006nz}, tunneling magnetoresistance,\cite{Tanaka:2001fj,Chun:2002rx} and current-driven magnetization control.\cite{Yamanouchi:2006pf}

Here, we report the observation of the current-perpendicular-to-the-plane (CPP) spin valve effect in ``self-exchange biased" bilayers wherein a magnetically soft FMS (GaMnAs) is exchange coupled with a magnetically hard metallic ferromagnet (MnAs). These materials have very different magnetocrystalline anisotropy constants ($K$) and Curie temperatures (\tc). For instance, at cryogenic temperatures, $K_{\rm{GaMnAs}} \sim 5$ kJ/m$^3$,\cite{Liu:2005nz} while $K_{\rm{MnAs}} \sim 500$ kJ/m$^3$.\cite{Ploog_PSS} Further, the distinct Curie temperatures (\tc $= 320$ K for MnAs, \tc $\lesssim 170$ K for (Ga,Mn)As) could be exploited to systematically tune the anisotropy and magnetization of the soft FM layer relative to the hard layer, hence providing a new model system for fundamental studies of exchange biasing and exchange spring behavior in interfacially coupled ferromagnets.\cite{GOTO:1965fk,Guo:2002qy,Ke:2004lr} The bilayer system introduced in this work may also provide new insights into spin-dependent scattering through a Bloch wall,\cite{Levy:1997lr} because the CPP geometry used in our experiments avoids complications introduced by the anisotropic magnetoresistance in the current-in-plane geometry.\cite{mibu:6442}  

All the heterostructures in this study are fabricated by molecular beam epitaxy on p-doped (001) GaAs substrates, after first depositing a 170 nm thick Be-doped GaAs buffer layer (with $p \sim 1 \times 10^{19} \rm{cm}^{-3}$) at a substrate temperature of 580 $^{\circ} \rm{C}$. We focus on two different sets of samples. Series A consists of five bilayer samples that contain 10 nm ``type-A'' MnAs on top of \gma~ ($x \sim 0.06$, thickness \tgma = 15, 30, 50, 80, 120 nm). In this particular crystalline orientation, the c-axis of the MnAs layer lies in the plane of the (001) GaAs substrate and the MnAs growth axis is $[\bar{1} 1 0 0]$.\cite{Ploog_PSS}  Growth conditions for the (Ga,Mn)As and MnAs layers are similar to those described elsewhere.\cite{Chun:2002rx} Series B consists of five MnAs/p-GaAs/(Ga,Mn)As trilayers with a fixed \gma~ layer thickness (30 nm) and composition ($x \sim 0.06$), and with varying thickness of the p-GaAs spacer layer (\tspacer =  1nm, 2 nm, 4 nm, 6 nm, and 8 nm). The p-doping in the spacer layer is nominally designed to be similar to that in the buffer layer.

Magnetic properties are characterized using a Quantum Design superconducting quantum interference device (SQUID) magnetometer. For magnetoresistance measurements, the samples grown on p-GaAs substrates are patterned into cylindrical mesas of 100 $\mu$m diameter using conventional photolithography and wet etching down to the p-substrate. After depositing insulating SiO$_{2}$ on the patterned mesas, Ohmic contacts (5 nm Ti + 100 nm Au) are evaporated through a window on top of the mesa.  The bottom (side) contact is formed by evaporating Ti/Au directly onto the etched substrates.  Magnetoresistance measurements are carried out in a current-perpendicular to the plane (CPP) geometry using a DC current source and digital multimeters. A pseudo-four-probe scheme is used with one current lead and one voltage probe on top of the mesa and separated current and voltage probes on the bottom contact. Since the resistance of the mesa is small ($\sim$300 m$\Omega$), a current of 4 mA is used to achieve good signal-to-noise ratio and avoid sample heating. We have checked that the current-voltage characteristics are linear during all our measurements. Finally, we note that in all the magnetization and magnetoresistance measurements discussed here, the external magnetic field is applied along [110] direction of the GaAs substrate. This corresponds to the $[1 1 \bar{2} 0]$ direction of the MnAs layer and is the easy axis for ``type-A" MnAs.

We first discuss the temperature- and magnetic field-dependent magnetization in MnAs/(Ga,Mn)As bilayers from series A. Figure 1 (a) shows the temperature dependent remanent magnetization $M(T)$ in three bilayer samples with \tgma = 30 nm, 50 nm and 120 nm, measured in a field of 30 Oe. The data clearly reveal the two different Curie temperatures for the MnAs (\tc = 318 K) and (Ga,Mn)As (\tc = 75 K) layers. Major magnetization hysteresis loops are measured after first saturating the MnAs layer in a 20 kOe field. An example is shown in Fig. 1 (b) for the sample with \tgma = 50 nm, revealing signatures of the very different coercive fields for MnAs (\hc $= 2$kOe) and (Ga,Mn)As (\hc $\sim 100$ Oe). Similar data are obtained for all the bilayers with \tgma $\geq 30$ nm. However, SQUID measurements on the sample with \tgma = 15 nm are unable to resolve the temperature- and field-dependent magnetization contribution from the (Ga,Mn)As layer. 

\begin{figure}[h]
\includegraphics[scale=0.8]{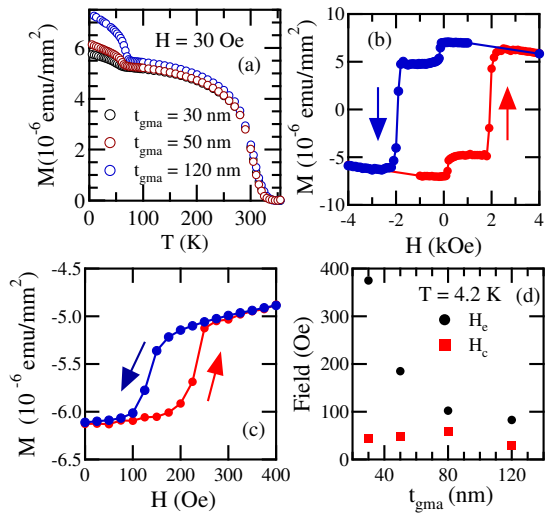}
\caption{(a) Temperature-dependent magnetization $M(T)$ for bilayer samples with three different (Ga,Mn)As thicknesses. Measurements are carried out while warming, after first cooling the samples down to 4.2 K in a field of 20 kOe. The magnetization is shown per unit area. (b) Major magnetization loop $M(H)$ at $T = 4.2$ K for a bilayer sample with \tgma = 50 nm, showing two distinct coercive fields for the two FM layers. (c) Minor magnetization loop for the same sample at $T = 4.2$ K, measured after saturating the MnAs layer in a field $H = -20$ kOe.(d) Dependence of \he (circles) and \hc (squares) on \tgma~ at $T = 4.2$ K.}
\end{figure}

Figure 1 (c) shows the minor hysteresis loop for the same sample as in Fig. 1 (b); these data are measured over a field range $-1$ kOe $\leq H \leq +1$ kOe, after first saturating the MnAs layer in a $- 20$ kOe field. The positive displacement of the center of the hysteresis loop from zero field is unambiguous evidence for a FM exchange coupling between the (Ga,Mn)As layer and the saturated MnAs layer. In other words, when the MnAs magnetization points along the negative field direction, the (Ga,Mn)As layer experiences an  exchange field (\he) along the direction of the saturated MnAs moment, resulting in a ``negative exchange bias." We confirm this by carrying out a similar measurement after saturating the MnAs layer in the direction of positive field and find that the minor hysteresis loop is shifted in the negative direction (data not shown). Similar behavior is observed in all the bilayer samples with \tgma $\geq 30$ nm. In Fig. 1(d), we show that \he~ decreases monotonically with the thickness \tgma~ of the (Ga,Mn)As layer. We find that \he $\sim  \frac{1}{t_{\rm{gma}}}$, qualitatively consistent with an analytical model for exchange-coupled FM/FM bilayers wherein a partial domain wall forms in the softer side of the interface.\cite{Guo:2002qy} Our data are inconsistent with the inverse square prediction of an earlier idealized model for an exchange coupled FM/FM bilayer.\cite{GOTO:1965fk} Figure 1 (d) also shows that the coercive field (\hc) of the (Ga,Mn)As layer does not vary much with \tgma. This contrasts with observations in exchange biased FM/AFM bilayers where a larger \he~ is typically accompanied by a larger \hc.\cite{Nogues:1999uq,Eid:2004rx} 

\begin{figure}[h]
\includegraphics[scale=0.8]{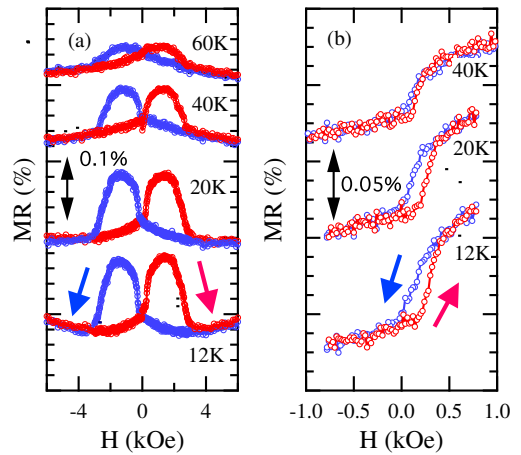}
\caption{Normalized CPP magnetoresistance (as a percentage) in a bilayer sample (\tgma = 30nm) in (a) major and (b) minor hysteresis loops at different temperatures. The data are offset vertically for clarity.}
\end{figure}

We now correlate our magnetization measurements with CPP magnetoresistance. Figure 2(a) shows the relative magnetoresistance ($(R (H) - R(0))/R(0)$) during a major hysteresis loop in a bilayer sample with \tgma = 30 nm. A comparison between the transport data and magnetization hysteresis loops provides evidence for the spin valve effect, showing a high (low) resistance state when the (Ga,Mn)As and MnAs layers have antiparallel (parallel) magnetization. As expected, the magnitude of the spin valve effect decreases with increasing temperature and eventually disappears above the \tc~ of (Ga,Mn)As. Figure 2(b) shows the exchange biased spin valve effect when the external field is swept over a range corresponding to a minor hysteresis loop: again, the effect decreases with increasing temperature, but interestingly reveals {\it reversible} behavior well before reaching \tc, even though the exchange biasing persists up to \tc. This transition to reversible behavior at higher temperatures is also observed in SQUID measurements of the field-dependent magnetization (data not shown). We speculate that -- at higher temperatures -- the coercivity of the (Ga,Mn)As becomes small enough to allow for ``exchange spring behavior'':\cite{Kneller_Hawig} in this temperature range, the (Ga,Mn)As layer shows no remanence because its magnetization adiabatically follows the net field comprised of the external field and the local exchange field. 

Finally, we measure the magnetization and CPP magnetoresistance in trilayer samples from series B in order to examine whether the interfacial exchange coupling can be mediated by a spacer layer. We note that previous studies of \gma/AlAs/MnAs tunneling structures have suggested a weak coupling even in the presence of a tunnel barrier.\cite{Chun:2002rx} Figure 3 shows both magnetization and CPP magnetoresistance measurements in a trilayer sample with a \tspacer = 2 nm, confirming that the exchange coupling is mediated by holes in the spacer layer and that an exchange biased spin valve effect persists in such devices. Similar data are observed in all six trilayer samples studied, and we find that \he~ decreases with increasing spatial separation between the FM layers, dropping off rapidly for \tspacer $\gtrsim 6$ nm  (data not shown). We caution though that a detailed understanding of the variation of \he~ with \tspacer~ will need to await more systematic measurements (e.g. using wedged spacer samples\cite{Tanaka:2001fj}) that minimize unavoidable growth-to-growth variations.

 \begin{figure}[h]
\includegraphics[scale=0.8]{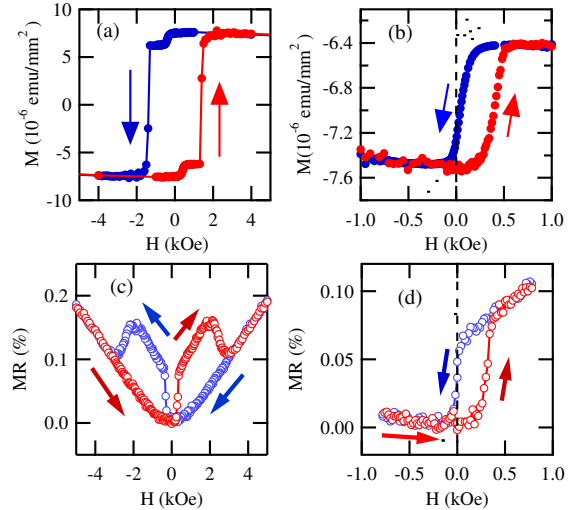}
\caption{(a) Major and (b) minor hysteresis loops for a trilayer sample (\tspacer = 2 nm) at 4.2 K, showing that exchange biasing persists even with a spacer layer. Panels (c) and (d) show the CPP magnetoresistance in major and minor loops, respectively, showing the spin valve effect in the trilayer.}
\end{figure}

In summary, we have demonstrated the CPP spin valve effect in self exchange biased ferromagnetic MnAs/(Ga,Mn)As bilayers, as well as in MnAs/p-GaAs/(Ga,Mn)As trilayers. The CPP spin valve effect in these systems likely arises from spin-dependent scattering at the interfaces but a detailed knowledge of the band structure, interfaces and impurity scattering is required for a quantitative understanding. We caution though that despite the observation of linear I-V characteristics, it is difficult to completely rule out the possibility of a tunneling-based magnetoresistance due to a small interfacial potential barrier. Nonetheless, the results reported here bring to light a new model system that can help address fundamental questions about the nature of spin-dependent transport and scattering at FM metal/semiconductor interfaces. Furthermore, the observation of exchange coupling between a metallic and semiconducting ferromagnet suggests novel approaches for the epitaxial and lithographic engineering of magnetic properties such as the coercivity and the Curie temperature of FMS such as (Ga,Mn)As. Finally, we note that the exchange biasing and spin valve effect behavior reported here also appear to occur in recent studies of hybrid trilayer structures that use phase-separated GaAs:MnAs instead of single phase (Ga,Mn)As as the FMS layer.\cite{hai_arXiv:0708.1681}
  
This research has been supported by grant numbers ONR N0014-05-1-0107, and NSF DMR-0305238 and -0401486. This work was performed in part at the Penn State Nanofabrication Facility, a member of the NSF National Nanofabrication Users Network.





\end{document}